\documentclass[fleqn,usenatbib]{mnras}

\usepackage{txfonts}

\usepackage[T1]{fontenc} 
\usepackage{ae,aecompl}

\usepackage[british]{babel}
\usepackage{color}
\usepackage{amssymb}
\usepackage{epsf,multirow}
\usepackage{graphicx}

\def\aap{A\&A}
\def\apj{ApJ}
\def\apjs{ApJS}
\def\apjl{ApJL}

\def\mnras{MNRAS}
\def\aj{AJ}
\def\nat{Nature}
\def\aaps{A\&A Supp.}
\def\prd{Phys. Rev. D}

\def\araa{ARA\&A}   
\def\aj{AJ}
\def\araa{ARA\&A}
\def\apj{ApJ}
\def\apjl{ApJ}
\def\apjs{ApJS}
\def\ao{Appl.Optics}

\def\aap{A\&A}

\def\aaps{A\&AS}

\def\mnras{MNRAS}

\def\prd{Phys.Rev.D}

\def\pasp{PASP}

\def\aj{AJ}
\def\araa{ARA\&A}
\def\apj{ApJ}
\def\apjl{ApJ}
\def\apjs{ApJS}
\def\ao{Appl.Optics}

\def\aap{A\&A}

\def\aaps{A\&AS}

\def\mnras{MNRAS}

\def\prd{Phys.Rev.D}

\def\pasp{PASP}

\defcitealias{Binney1982}{BM82}
\newcommand{\citeta}{\citetalias}

\def\Kb{$K$-band}
\def\Vb{$V$-band}
\def\Rb{$R$-band}
\def\kms{~km~s$^{-1}$}
\def\micron{~$\mu$m}

\def\deg{$^{\circ}$}

\newcommand{\Ls}{L$_{\sun}$}
\newcommand{\msun}{M_{\sun}}

\def\n{NGC~}
\def\S{SINFONI}
\def\HST{{\it HST}}

\def\Ms{M$_{\sun}$}
\def\Mbh{M$_{\rm BH}$}
\def\mbh{M_\mathrm{BH}}

\def\ml{\Upsilon}

\def\H2{H$_2$}

\def\Hb{H$\beta$}

\def\NII{[N{\sc ii}]}

\title[SMBH and nuclei of \n5419]{The supermassive black hole and
  double nucleus of the core elliptical \n5419\thanks{Based on observations at the
    European Southern Observatory (ESO) Very Large Telescope
    [083.B-0126(B)].}}

\author[Mazzalay et al.]{Ximena Mazzalay,$^{1}$\thanks{E-mail: ximena@mpe.mpg.de} Jens Thomas,$^{1,2}$  Roberto P. Saglia,$^{1,2}$ Gary A. Wegner,$^{3}$
\newauthor Ralf Bender,$^{1,2}$ Peter Erwin,$^{1}$ Maximilian H. Fabricius,$^{1,2,4}$ Stephanie P. Rusli$^{1,2}$\\ 
$^{1}$Max-Planck-Institut f\"ur extraterrestrische Physik, Postfach 1312, 85741 Garching, Germany\\ $^2$Universit\"atssternwarte, Scheinerstrasse 1, 81679 M\"unchen, Germany\\ 
$^3$Department of Physics and Astronomy, Dartmouth College, 6127 Wilder Laboratory, Hanover, NH 03755, USA\\
$^4$Subaru Telescope, 650 North Aohoku Place, Hilo, HI 96720}

\date{Accepted XXX. Received YYY; in original form ZZZ}

\pubyear{2015}

\begin{document}
\label{firstpage}
\pagerange{\pageref{firstpage}--\pageref{lastpage}}
\maketitle

\begin{abstract}

  We obtained adaptive-optics assisted \S\ observations of the central
  regions of the giant elliptical galaxy \n5419 with a spatial
  resolution of $0.2$~arcsec ($\approx 55$~pc). \n5419 has a large
  depleted stellar core with a radius of 1.58~arcsec (430~pc). \HST\
  and \S\ images show a point source located at the galaxy's
  photocentre, which is likely associated with the low-luminosity AGN
  previously detected in \n5419.  Both the \HST\ and \S\ images also
  show a second nucleus, off-centred by 0.25~arcsec ($\approx
  70$~pc). Outside of the central double nucleus, we measure an almost
  constant velocity dispersion of $\sigma \sim 350$\kms.  In the
  region where the double nucleus is located, the dispersion rises
  steeply to a peak value of $\sim 420$\kms. In addition to the \S\
  data, we also obtained stellar kinematics at larger radii from the
  South African Large Telescope. While \n5419 shows low rotation ($v <
  50$\kms), the central regions (inside $\sim 4 \, r_b$) clearly
  rotate in the opposite direction to the galaxy's outer parts.  We
  use orbit-based dynamical models to measure the black hole mass of
  \n5419 from the kinematical data outside of the double nuclear
  structure. The models imply $\mbh=7.2^{+2.7}_{-1.9} \times
  10^9$~\Ms. The enhanced velocity dispersion in the region of the
    double nucleus suggests that \n5419 possibly hosts two
    supermassive black holes at its centre, separated by only $\approx
    70$~pc.  Yet our measured $\mbh$ is consistent with the black
    hole mass expected from the size of the galaxy's depleted stellar
    core. This suggests, that systematic uncertainties in $\mbh$
    related to the secondary nucleus are small.
\end{abstract}



\section{Introduction}\label{s_intro}

It is now well established that super massive black holes (SMBHs) are
present at the centres of galaxies with bulges. The observational
correlations found between the black-hole mass and various parameters
of the host galaxy, e.g., velocity dispersion, bulge mass, and bulge
luminosity, have led to the idea that the formation and evolution of
early type galaxies and their nuclear SMBHs are tightly linked
\citep[see e.g.][and references therein]{Kormendy2013}.

\n5419 is a luminous elliptical galaxy ($M_V = -23.1$). Such bright
early type galaxies (brighter than $M_V \sim -21$) often have a low
density `core' with a typical size of a few tens or hundreds of
parsecs. Inside the core or break radius, $r_b$, the light profile is
much shallower than the inward extrapolation of the outer S\'ersic
profile \citep[e.g.][]{Lauer1995,Lauer2005,Graham2003a,Rusli2013b}.
Moreover, core galaxies differ from fainter ellipticals in their
isophotal shapes and degree of rotational support: core ellipticals
have boxy instead of disky isophotes and are supported by anisotropic
velocity dispersions rather than rotational stellar motions
\citep[e.g.][]{Nieto1991a,Kormendy1996a,Faber1997,Lauer2012a}.  These
morphological and kinematic distinctions between core galaxies and
fainter ellipticals suggest that the processes involved in their
formation are different.

The most plausible mechanism for core formation is black hole scouring
that occurs in non-dissipative mergers of galaxies as a result
of the dynamics of their central SMBHs: the SMBHs spiral into the
centre of the merger via dynamical friction, ultimately forming a
black hole binary \citep[][]{Begelman1980}. As the binary shrinks, it
ejects stars on intersecting orbits, creating a low-density core with
a mass deficit of the order of or a few times larger than the mass of
the binary \citep[see e.g.][and references
therein]{Milosavljevic2001a,Merritt2006a,Merritt2013a}. The black hole
binary model can explain the correlations between core structure,
black hole masses, mass deficits and the observed orbital structure in
core galaxies \citep[e.g.][]{Faber1997,Milosavljevic2001a,
  Kormendy2009a, Hopkins2009a,Thomas2014,Thomas2016}. Dissipation-driven core
formation has been reported from numerical $N$-body simulations that
include the dynamical effects of AGN feedback, but these results have
not been tested yet in detail against the wealth of observations
available for core elliptical galaxies.

Core scouring implies the formation of black hole binaries.  Their
subsequent evolution can follow different paths.  In particular, if
the binary separation decreases enough and the SMBHs manage to
coalesce, the merged black hole could be ejected from the galaxy
centre by anisotropic emission of gravitational waves \citep[e.g.][]{Begelman1980}.  
As the probability that the remnant black hole recoils at a velocity
exceeding the escape velocity of a large elliptical galaxy is low
\citep[e.g.][]{Lousto2012a}, it will most likely remain bound to the
galaxy on a radial orbit.  On the other hand, if the binary decay
stalls, subsequent mergers may bring a third SMBH or second black hole
binary to the centre \citep[e.g.][]{Valtonen1996}. The interactions of
this newly formed multiple SMBH system will eventually displace one or
more of the SMBHs from the nucleus.  In both of these scenarios,
off-centred SMBHs are expected to be found in the centre of bright
elliptical galaxies. The observational evidence of binary or recoiled
SMBHs that supports these theoretical predictions is still scarce,
usually involving pairs of SMBHs at kiloparsec separations 
\citep[e.g.][]{Komossa2003a,Ballo2004,Bianchi2008,Koss2011,Fu2011a,McGurk2015,Comerford2015}. On
smaller scales, mostly indirect evidence has been reported, with only
one secure case known so far \citep[CSO 0402$+$379,
][]{Rodriguez2006}.

In this paper we study \n5419, the dominant galaxy of the poor cluster
Abell~753.  This large elliptical galaxy was first classified as a
core by \citet{Lauer2005}, who modelled its inner $\sim 10$~arcsec
surface brightness profile derived from a {\it HST}/WFPC2 F555W
image. 
Based on a Nuker-law \citep{Lauer1995} fit to \n5419 profile, they
derived a break radius $r_b = 2.38$~arcsec (650~pc), making this relatively
low-surface brightness core the largest among the 42 objects
classified as core galaxies in their sample.  Moreover, the {\it HST}
image presented by \citet[][]{Lauer2005} revealed the presence of a
double nucleus at the centre of the galaxy, with a projected
separation of a few tens of parsecs (see also \citealt{Capetti2005b,Lena2014}; this work). 
Additionally, radio observations
\citep[][]{Goss1987,Subrahmanyan2003} and the detection of hard X-rays
\citep[][]{Balmaverde2006} indicate that the centre of \n5419 hosts a
low-luminosity AGN (LLAGN). All this makes \n5419 an interesting case for studying  
the interplay between core galaxies and their SMBHs. 

We observed \n5419 as part of our black hole survey, consisting of 30 galaxies observed with \S\ at the Very Large Telescope (VLT, \citealt{Nowak2007, Nowak2008, Nowak2010, Rusli2011, Rusli2013a, Rusli2013b, Saglia2016}; Bender et al. in preparation; Erwin et al. in preparation; Thomas et al. in preparation).  
Here we report the results obtained for \n5419. The \S\ data of \n5419 and additional long-slit spectroscopy from the Southern African Large Telescope are presented in Section~\ref{s_data}. In Sections~\ref{s_photo} and \ref{s_kin} we present the results from the imaging and spectroscopy of \n5419. Section~\ref{s_dyn} deals with the dynamical modelling of the galaxy. A discussion about the double nucleus of \n5419 can be found in Section~\ref{s_nucleus} and our final conclusions in Section~\ref{s_summary}. 

We assume a distance to \n5419 of 56.2~Mpc, derived from the radial velocity corrected for Local Group infall onto Virgo taken from Hyperleda, $v_{\rm vir}=4047$\kms, and a value of $H_0 = 72~{\rm km~s^{-1}~Mpc^{-1}}$. At this distance, 1~arcsec corresponds to 273~pc.

\section{Observations and data reduction}\label{s_data}

\subsection{\S\ IFS data}\label{s_SINFdata}
Adaptive-optics-assisted near-infrared (NIR) observations of \n5419 were obtained with the \S\ integral field spectrograph at the 8~m VLT UT4 on 2009 May 20$-$24. A field of view (FOV) of $\sim 3 \times 3$~arcsec ($820 \times 820$~pc) was covered at a spatial sampling of $0.05 \times 0.10$~arcsec$^2$~pixel$^{-1}$ in the \Kb\ (1.95$-$2.45\micron) with a spectral resolution of ${\rm R}\sim 5000$. 
The observations were performed according to a standard object-sky-object strategy, followed by the observation of a point spread function (PSF) star to assess the adaptive optics (AO) performance and characterize the PSF  \citep[see e.g.][]{Rusli2013a,Mazzalay2013a}. 
Individual 10~min exposures dithered by a few spatial pixels resulted in a total on-source exposure time of 250~min. 
During the observations, the AO was operated in the laser guide star mode. This improves the spatial resolution to $\approx 0.2$~arcsec (55~pc), as given by the full width at half maximum (FWHM) of the PSF stars associated with the observations.

The \S\ data were reduced using {\sc esorex} \citep{Modigliani2007}, including all the standard reduction steps, i.e. bias subtraction, flat-fielding, bad pixel removal, detector distortion and wavelength calibration, sky subtraction \citep{Davies2007a}, reconstruction of the object data cubes and telluric correction. The data cubes were re-sampled to a spatial scale of $0.05 \times 0.05$~arcsec$^2$~pixel$^{-1}$. 25 data cubes were combined to produce one single final cube. The lower panel of Fig.~\ref{f_spectra} shows an example spectrum in the region of the NaI and CO absorption lines, obtained from the combination of the entire \S\ FOV. No emission lines are observed in our data. 

\subsection{Long-slit SALT data}\label{s_SALTdata}

In addition to the NIR data for the centre, optical spectra along the position angles\footnote{Position angles are measured from north to east.}  
PA=78\deg\ (hereafter MJ axis) and PA=168\deg\ (hereafter MN axis)
were obtained at the 10~m Southern African Large Telescope (SALT). The
observations utilized the Robert Stobie Spectrograph (RSS) in long
slit mode (slit width 1.25~arcsec) and grating PG2300 covering the
wavelength range 4800$-$5700~\AA. The plate scale is
0.14~arcsec~pix$^{-1}$ and the spectral resolution, measured from the
arc lamps, is ${\rm FWHM} \sim 40$\kms.  Three 850 second exposures
were obtained on 2013 Mar 22 for PA 78\deg\ and two 1100 second
exposures for PA 168\deg\ were made on 2013 Apr 30. The bias-subtracted and flattened images provided by the SALT pipeline were wavelength calibrated in two dimensions by fitting ThAr and CuAr
comparison spectra with fifth order polynomials employing the
long-slit menu in {\sc iraf}.  Flux calibrations used the G93-48
standard in {\sc iraf}.

\begin{figure}
\centering
\includegraphics[width=1.\columnwidth]{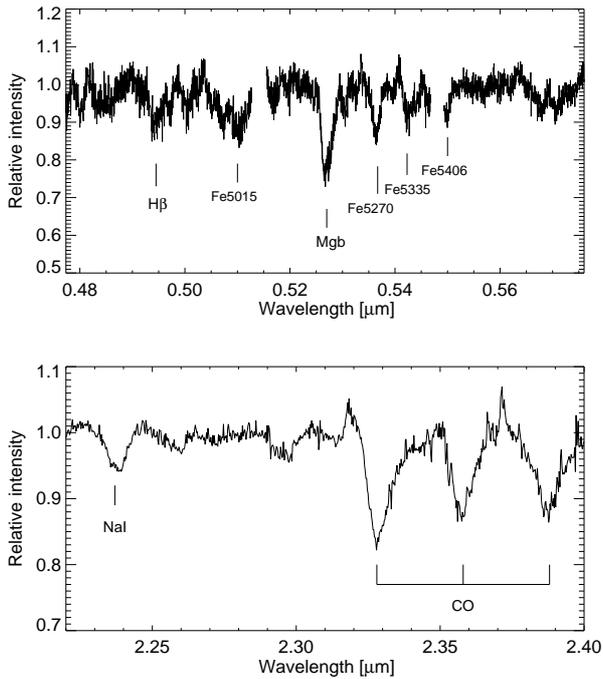}
\caption {Optical (SALT, upper panel) and NIR (\S, lower panel) continuum-normalized spectrum of \n5419, in the observed frame. The labels indicate the principal lines observed in the galaxy spectra.}
\label{f_spectra}
\end{figure}

The upper panel of Fig.~\ref{f_spectra} shows an example SALT spectrum in the observed frame, obtained by summing up the signal over three pixels (0.42~arcsec) around the continuum maximum, and continuum normalized. 

\section{The light profile: a depleted core with a double nucleus}\label{s_photo}

\begin{figure*}
\centering
\includegraphics[width=1.\textwidth]{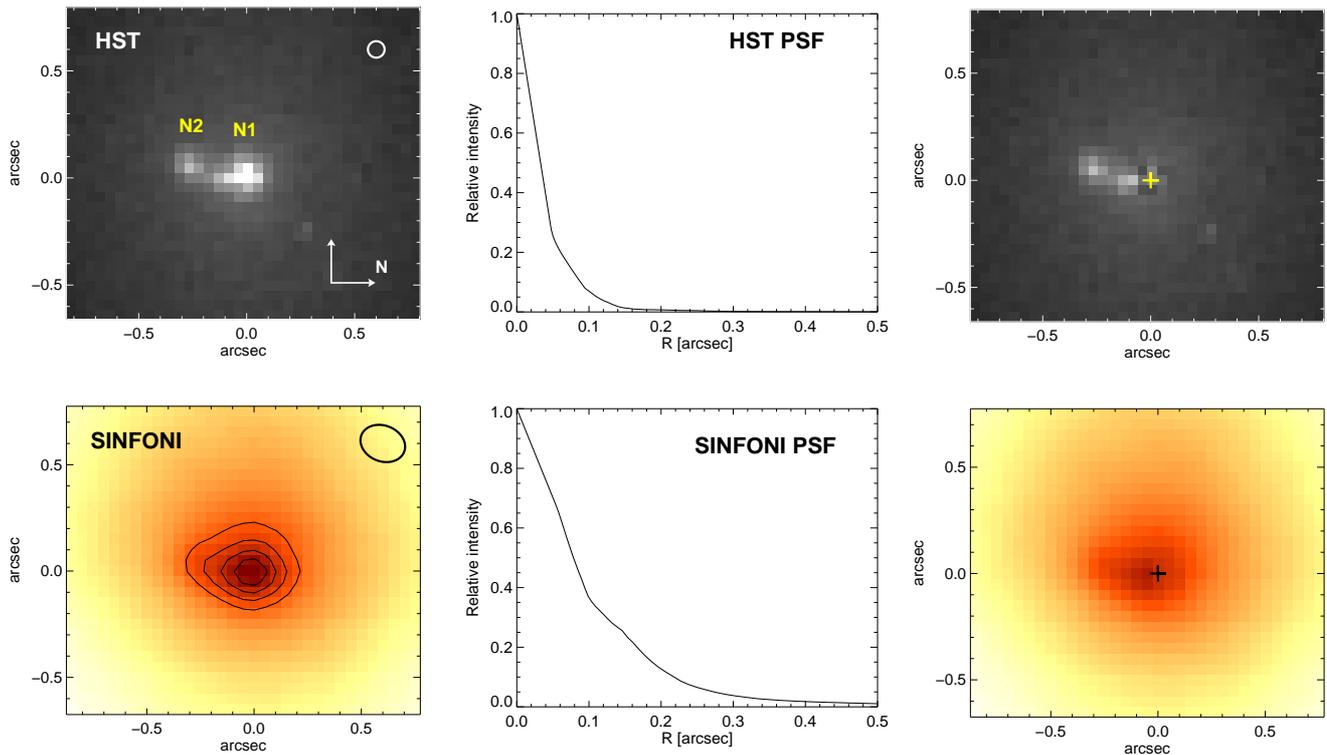}
\caption {Close-up of the inner $1.5 \times 1.5$~arcsec ($410 \times 410$~pc) of \n5419. 
Left panels: {\it HST}/WFPC2 {\it F555W} image and \Kb\ VLT/\S\ collapsed  cube and isophotes. The images are centred at the position of the continuum maximum; the spatial orientation is indicated in the upper panel. N1 and N2 denote the central point source and the off-centre nucleus, respectively. The ellipses on the upper-right corners of the images indicate the Gaussian FWHM of the respective PSFs. 
Central and right panels show the corresponding PSF profiles and point-source subtracted images.}
\label{f_ims}
\end{figure*}

\begin{figure}
\centering
\includegraphics[width=0.95\columnwidth]{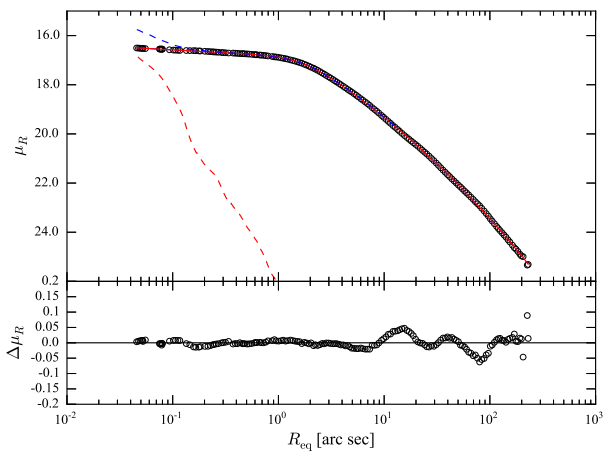}
\caption{Surface brightness profile (circles), best-fitting Core-S\'ersic model (red-solid line) and residuals. The blue- and red-dashed curves correspond to the $HST$ surface brightness profile of \n5419 before the point source subtraction and the rescaled $HST$ PSF profile, respectively (see Section~\ref{ss_core} for details).}
\label{f_corefit}
\end{figure}

To study the light distribution of \n5419 we use {\it HST}/WFPC2 F555W\footnote{proposal ID $=$ 6587, PI $=$ D. Richstone} and 3.6\micron\ {\it Spitzer}
IRAC1\footnote{program ID $=$ 30318, PI $=$ G. Fazio} images retrieved
from the archives, as well as our \S\ collapsed cube. 
Overall, \n5419 shows a smooth light distribution, with no signs of obvious distortions that
could indicate a recent merger or the presence of dust.  
The {\it HST} image shows what appears to be a double nucleus at the centre of the galaxy (upper-left panel of Fig.~\ref{f_ims}), first seen by \citet{Lauer2005} and \citet{Capetti2005b}. 
The brighter of the two nuclei, N1, is located at the galaxy's photocentre\footnote{A low-significance displacement of 0.6~pixel (0.02~arcsec) between the central point-source and the galaxy's photocentre of \n5419 is reported by \citet{Lena2014}.}. 
It is unresolved in the {\it HST} image and is likely associated with the LLAGN of \n5419. 
The second nucleus, N2, is seen in the form of a bright blob, off-centred by approximately 0.25~arcsec (70~pc) towards the south, almost aligned with the semi-minor axis of the galaxy. The off-centre nucleus has a Gaussian FWHM of $\approx 0.15$~arcsec, slightly larger than the {\it HST} PSF ($0.07$~arcsec). 
Additionally, a `bridge' of weak emission is observed between the two nuclei, suggesting that they are physically related and that their proximity is not merely a projection effect.

\subsection{The depleted stellar core in \n5419}\label{ss_core}

The surface brightness profile of \n5419 was constructed by fitting ellipses to 
the {\it HST} and {\it Spitzer} images. For this we employed the task ellipse of the
{\sc stsdas} package of {\sc iraf}. 
As a first step, we used a rescaled $HST$ PSF to subtract the central point source N1 from the $HST$ image.     
The scaling factor was computed iteratively to make the slope of the light profile inside 0.2~arcsec  consistent with the one around 1~arcsec. 
Once N1 was subtracted (upper-right panel of Fig.~\ref{f_ims}), we derived the final $HST$ profile masking N2, as well as several globular clusters.   
Finally, the {\it HST} and {\it Spitzer} profiles were matched by determining the scaling and sky
value that minimize the magnitude square differences between the two
profiles in the 5$-$18~arcsec range. Since the galaxy fills the entire
{\it HST} FOV, the sky background was estimated from the larger
{\it Spitzer} image. Fig.~\ref{f_corefit} shows the 
final surface brightness profile, with and without point source, calibrated to Cousins $R$-band using
the aperture photometry reported by \citet{Poulain1994}. The scaled $HST$ PSF is also included.  

The remaining light profile of \n5419 without the two compact central
sources is well described by a Core-S\'ersic function 
\citep[e.g.][]{Graham2003a, Trujillo2004a, Rusli2013b}. This is shown in Fig.~\ref{f_corefit}, 
where we plot the surface brightness profile together with the corresponding Core-S\'ersic fit
and its residuals. The best-fitting model has a S\'ersic index $n=7.2$,
a projected half-light radius $r_e=110.8$~arcsec, a core break radius
$r_b=1.58$~arcsec, a surface brightness at the break radius of
$\mu_b=17.09$~mag~arcsec$^2$, a flat inner slope $\gamma=0.09$ and a
transition parameter $\alpha=3.22$.   
Note that the Core-S\'ersic fit 
provides a much better description of \n5419's light profile than 
a simple S\'ersic model, and clearly identifies this galaxy as 
a core elliptical (e.g. \citealt{Rusli2013b}). 
Given these parameters, we can estimate the light deficit in the core from
the luminosity difference between the best-fitting Core-S\'ersic model
and its S\'ersic component. We follow \citet{Rusli2013b} and obtain a
$R$-band luminosity deficit of $(3.78\pm0.15) \times 10^9 $~\Ls\
(assuming an extinction correction of $A_R = 0.193$) . This
corresponds to $M_{R}=-19.5$~mag or $M_{V}=-18.8$~mag (with $V-R=0.67$ from \citealt[][]{Poulain1994}).

\subsection{The double nucleus as seen by \S}
\label{ss_doublenuc}

\begin{figure}
\centering
\includegraphics[width=.8\columnwidth]{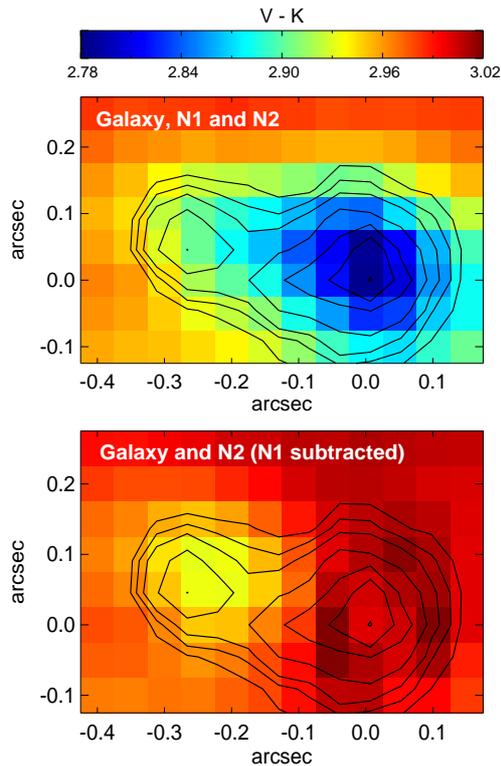}
\caption {$V-K$ colour maps derived form the {\it HST} and \S\ images before (upper panel) and after (lower panel) the subtraction of the central point source. The contours correspond to the {\it HST} isophotes (see Fig.~\ref{f_ims}). North is right, east is up.}
\label{f_color}
\end{figure}

The double nucleus mentioned above has only been detected in the optical
with {\it HST} thus far. Although they are not resolved, these two nuclei are also seen in our \S\ data.
The lower-left panel of Fig.~\ref{f_ims} shows a zoom into the central $r\sim 0.5$~arcsec (135~pc) of the \S\ image 
of \n5419 centred at the continuum maximum. The \Kb\ isophotes are overlaid. Taking
into account the differences in spatial resolution between the \HST\
and \S\ images, it is clear that the elongation of the isophotes near
the centre is the result of the presence of the secondary nucleus. 

We used these images to derive the $V-K$ colour of the different
components in the centre of \n5419.  We calibrated the \HST\ image to
the \Vb\ in the Johnson system using the transformation from
\citet{Holtzman1995a} and calibrated the \S\ collapsed cube to the
\Kb\ using a 2MASS image. 
We degraded the spatial resolution and pixel scale of the \HST\ image to match 
those of our \S\ data and created a $V-K$ colour map. 
The upper panel of Fig.~\ref{f_color} shows a close-up around N1 and N2 of 
the resulting $V-K$ colour map. 
The contours correspond to the \HST\ isophotes. 
The central point source stands out from the rest of the galaxy by its bluer colour. 
While the galaxy colour, measured in regions further away from N1 and N2, varies from $\approx 2.99-3.02$, 
N1 has a colour $V-K \approx 2.8$. 
However, this value does not represent that of N1 alone, since it also includes light from the galaxy. 
Similarly as for the \HST\ image, we subtracted the central point source from the \S\ image (lower-right panel of Fig.~\ref{f_ims}) and derived the intrinsic $V-K$ colour of N1 from the integrated 
magnitudes of the \HST\ and \S\ rescaled PSFs. Unlike in the optical, N1 is not prominent in the \Kb\ 
(it accounts for only $\sim 5$~percent of the light coming from the inner $r=0.15$~arcsec), 
resulting in a relative blue colour of $V-K =1.13$. 

The lower-panel of  Fig.~\ref{f_color} shows again the $V-K$ colour map but this time with 
the central point source subtracted. 
A decrease of the $V-K$ colour can be clearly seen around the region where the secondary nucleus is located. 
To obtain its intrinsic colour we isolated the light from N2 by subtracting 
from the N1-subtracted \HST\ and \S\ images a galaxy model constructed from the ellipse-fitting results and measured the magnitudes in an aperture of $r=0.15$~arcsec ($40$~pc) centred at N2. The absolute magnitudes of N2 in the $V$- and $K$-bands are
${M_V}=-11.93$ and ${M_K}=-13.61$, corresponding to luminosities of
${L_V}=4.9 \times 10^6$~\Ls\ and ${L_K}=5.7 \times 10^6$~\Ls. The
intrinsic colour of N2 is thus $V-K=1.68$. 
While the colour of \n5419 is typical of an old early type galaxy \citep[e.g.][]{Bower1992}, the off-centre 
nucleus would correspond to a younger stellar population. 

A summary of the measured luminosities and colours of the central point source and the off-centre nucleus are given in Table~\ref{t_color}. All values have been corrected for Galactic extinction ($A_V=0.199$ and $A_K=0.022$ from NED).

\begin{table}
\caption{Luminosities and colour of the central point source N1 and the off-centre nucleus N2 measured inside apertures of 0.15~arcsec in radius. The $V-K$ colour of the galaxy is also included for comparison.}
\label{t_color}
\centering
\begin{tabular}{cccc}
\hline
  & $L_V$ [\Ls] & $L_K$ [\Ls] & $V-K$ \\
\hline
N1 & $1.3 \times 10^7$ & $9.2 \times 10^6$ & 1.13\\
N2 & $4.9 \times 10^6$ & $5.7 \times 10^6$ & 1.68\\
Galaxy & -- & -- & $\sim 3$\\
\hline
\end{tabular}
\end{table}

\subsection{Stellar mass of the off-centre nucleus}\label{ss_massnuc}
  
Assuming that the luminosity of N2 is entirely of stellar origin, we use the
single stellar population (SSP) models of
\citet{Maraston1998,Maraston2005} to derive some of its basic 
properties. For a Salpeter initial mass function (IMF), the colour of N2 
corresponds to a metal-poor, $\sim 2$~Gyr old stellar population with a \Vb\ 
mass-to-light ratio M/L$_{V} \simeq 1$. This M/L$_{V}$ ratio together
with the derived luminosity implies a stellar mass for the off-centre
nucleus of about $5 \times 10^6$~\Ms\ inside a region of $\approx 40$~pc
of radius. 

Our stellar mass estimate lies in the range of extremely massive
clusters and low-mass dwarf galaxies and ultra-compact dwarfs
\citep[e.g.][]{Walcher2005,Erwin2015a}.  Given the blue colour and
young age, it is unlikely that N2 is a massive globular
cluster.  Globular clusters in early-type galaxies are typically old
\citep[ages $\gtrsim 10$~Gyr; e.g.][]{Hempel2007a,Chies-Santos2011},
although younger clusters ($\sim 2-8$~Gyr) have been claimed in a few
cases \citep[e.g.][]{Goudfrooij2001,Puzia2002,Hempel2003}. Even an
intermediate-age cluster is unlikely, since they usually have a
$V-K>2$.  We can further compare the \Vb\ magnitude of N2 with
those of the other clusters observed in the \HST\ image of
\n5419. Their magnitude does not vary much from cluster to cluster,
with a mean value of $M_V= -10.45 \pm 0.60$~mag estimated from ten
clusters. These are relatively bright compared to, for example, the
ones found in the Milky Way or M31. Still, N2 is about an order
of magnitude brighter.  
Therefore it is unlikely that the off-centre nucleus is a globular cluster. 

It is possible that some fraction of the light we see is
due to AGN emission, and that the colour of the stellar population in
the secondary nucleus is in fact redder than assumed here. This would
imply a higher mass-to-light ratio. However, even in the most extreme
cases, the stellar mass would only increase by a factor of five.

\section{Kinematics and line strength measurements}\label{s_kin}

In order to obtain the kinematic information of \n5419 we used two
different datasets: \S\ integral field spectroscopic (IFS) data covering the inner regions of the
galaxy with high spatial resolution and long-slit SALT data obtained
along the MJ and MN axis providing information out to $\sim
15$~arcsec. In the following sections we describe the extraction of
the kinematics from these datasets and the results. Particular care
was taken in estimating the contribution from non-stellar emission to
the NIR continuum, which could bias the derived kinematics.
Additionally, in Section~\ref{s_indices} we measure optical line
indices to study the stellar populations.

\subsection{\S\ kinematics: a high dispersion region near the centre}\label{s_SINFkin}

\begin{figure*}
\centering
\includegraphics[width=1.\textwidth]{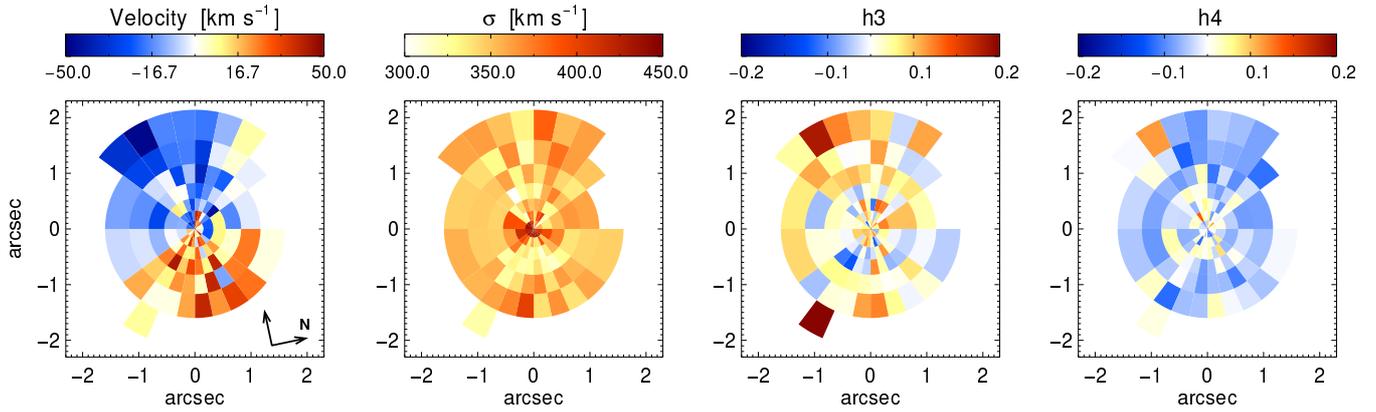}
\caption {Kinematic maps of \n5419 derived from the \S\ data. The spatial orientation is indicated in the left panel; the major axis of the galaxy is aligned with the $y$-axis.}
\label{f_sinf_kin}
\end{figure*}

\begin{figure}
\centering
\includegraphics[width=\columnwidth]{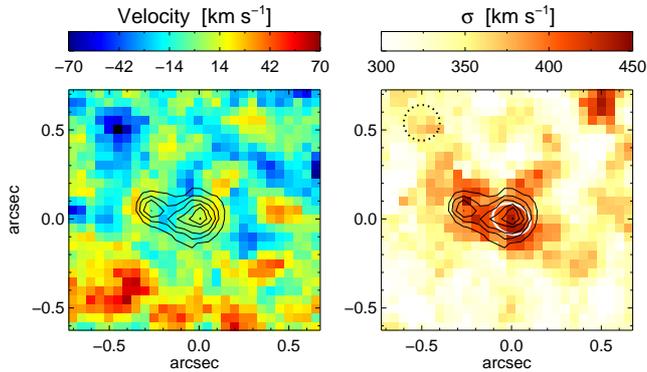}
\caption {Velocity and velocity dispersion maps of the inner region of
  \n5419 derived from the \S\ data. North is right and east is up. Overplotted
  are the isophotes of the {\it HST} image (see Fig.~\ref{f_ims}). The dashed circle and white ellipse in the right panel indicate the binning size and the Gaussian FWHM of the \S\ PSF, respectively.}
\label{f_sinf_kin_zoom}
\end{figure}

Non-parametric line-of-sight velocity distributions (LOSVDs) were derived using a maximum penalized likelihood technique \citep[MPL,][]{Gebhardt2000a}. We apply the approach described in detail by \citet{Nowak2007,Nowak2008}. In brief, a set of late type stellar template spectra (observed with \S\ in the same configuration as the galaxy) were convolved with the LOSVDs in order to match the continuum-subtracted galaxy spectra in the region of the first two CO bandheads [$^{12}$CO(2$-$0) and $^{12}$CO(3$-$1)]. Only stars with CO equivalent widths ($W_{\rm CO}$) similar to the galaxy were used in the fitting to minimize template mismatch. 
In order to have a relatively high SNR (at least $\sim 40$), the individual spectra of the galaxy were combined using a radial and angular binning scheme following \citet{Gebhardt2003}. 
The uncertainties in the derived LOSVDs were estimated from a set of 100 Monte Carlo simulations of the galaxy spectra. These were created by adding different amounts of noise to the best fitting stellar template convolved with the derived LOSVD. 

While the non-parametric LOSVDs are used in the dynamical modelling
(see Section~\ref{s_dyn}), it is illustrative to express the LOSVD in
Gauss-Hermite moments. Fig.~\ref{f_sinf_kin} shows the velocity,
velocity dispersion, $h_3$ and $h_4$ maps derived from the
parametrization of the LOSVDs by Gauss-Hermite expansions up to order
four. The terms $h_3$ and $h_4$ quantify the asymmetric and symmetric deviations 
of the LOSVDs from a Gaussian function, respectively. 
The \S\ kinematic maps show that \n5419 is dispersion dominated,
with a velocity dispersion around $\sim 350$\kms\ and rotational motions
with an amplitude of no more than $\sim 50$\kms. $h_4$ is predominantly
negative over the entire \S\ FOV.

The velocity dispersion map of Fig.~\ref{f_sinf_kin} shows a hint of
an increase in $\sigma$ in the innermost bins ($r \lessapprox
0.4$~arcsec), almost reaching 450\kms.  To explore this in more detail
we relaxed the constraint on the minimum required SNR defining our
binning scheme and derived the kinematics for the inner $r\sim 
0.5$~arcsec again. In order to preserve as much of spatial
information as possible and, at the same time, ensure a SNR high
enough to obtain meaningful kinematic parameters, we integrated the
spectra over a circular aperture of $r=2$~pixels at each spatial
position. Fig.~\ref{f_sinf_kin_zoom} shows the velocity and velocity
dispersion maps of the inner $r \sim 0.5$~arcsec obtained in this
way. Since the SNR of the spectrum at each pixel is relatively low, the
resulting maps are rather noisy.  However, there is an extended region
with a high velocity dispersion at the centre. It is elongated in the
N$-$S direction and seems associated to the two nuclei in the centre,
approximately following the isophotes derived from the {\it HST} image
(see Section~\ref{s_photo}). 
We note that the extent of this region is much larger than the \S\ PSF shown by the 
white ellipse. Unlike in the dispersion map, there is no
particular pattern in the rotational velocity in the innermost 100~pc
of \n5419.

\subsection{Is the non-stellar continuum affecting our kinematic measurements?}\label{s_WCO}

\n5419 is known to harbour a LLAGN at its centre.  
Many early-type galaxies contain compact, high-brightness temperature radio cores
associated with LLAGNs and, in a few cases, parsec-scale jets are also
observed \citep[see e.g.][and references therein]{Ho2008a}. 
While there is no evidence of a well-defined jet on scales of arcseconds or
higher in \n5419, this galaxy does have a compact radio core \citep[$r
< 0.7$~arcsec; e.g.][]{Goss1987,Subrahmanyan2003} similar to those typically found in LLAGNs.
It is possible, then, that the off-centre nuclear structure is 
a jet similar to the optical jet observed in M87.

Since we are interested in the kinematics of the stars, which is sensitive
to the equivalent width of the CO lines, it is important
to determine how much this quantity is altered by the presence of a
non-stellar continuum.  For a stellar population that contains
late-type stars, it has been shown that the $W_{\rm CO}$ is rather
independent of the star formation history and age
\citep[e.g.][]{Davies2007b}; any additional contribution to the NIR
from a non-stellar continuum will dilute the $W_{\rm CO}$.
In order to estimate the contribution from non-stellar emission to the
NIR continuum, we measured the equivalent width of the CO(2$-$0) line
at 2.29\micron\ in the unbinned \S\ spectra. Due to the low SNR of the individual
spectra, the resulting $W_{\rm CO}$ map is rather noisy. However, it does not show 
any particular global pattern or any sign of a gradient towards either of the two nuclei, 
suggesting that the non-stellar emission coming from the centre is not 
enough to significantly alter the equivalent width of the CO. 
The $W_{\rm CO}$ is consistent with a constant value of 15.7~\AA\ over the entire 
\S\ FOV, with a standard deviation of 2.2~\AA. We assume that this is the
intrinsic $W_{\rm CO}$ of \n5419. 

We can further compare this value with the ones measured from higher SNR spectra obtained by integrating our data in apertures of $r=0.15$~arcsec centred at the position of the central point source N1 and the off-centre nucleus N2, $W_{\rm CO}=14.5 \pm 1.5$ and $15.3 \pm 1.0$, respectively. 
The maximum change in equivalent width is measured at N1, with a decrease of less than 10 percent. This would be expected if the non-stellar continuum contributes about 10 percent of the light in the CO region. 
Note that these numbers are consistent with the fractional light associated with N1, as estimated from the PSF subtraction of the \S\ collapsed cube (see Section~\ref{s_photo}). 
Therefore, even if all the extra light at the galaxy's unresolved photocentre 
comes from an AGN, its contribution to the total light in the centre would be 
very small (at most 10 percent) and the change in the CO equivalent width would 
be negligible. We checked that a 10 percent AGN contribution to the CO region 
has 
no effects on the derived stellar kinematics. 

In summary, the amount of any non-stellar light in the centre of \n5419 is not enough to affect our kinematic measurements. Moreover, since the extended high-$\sigma$ region at $r<0.35$~arcsec ($\approx 100$~pc) is not an artefact related to uncorrected continuum emission, it is unlikely that the off-centre source is an optical jet. Additional evidence against the jet scenario comes from the comparison of the colour of N2 with that of the well-studied optical jet in M87. The mean $V-K$ value of M87's jet is close to 1 \citep{Zeilinger1993,Stiavelli1997}, much bluer than the value of 1.68 we obtained for N2 (Section~\ref{ss_doublenuc}). 

\subsection{Long-slit kinematics: a counter rotating core}\label{s_SALTkin}

\begin{figure}
\centering
\includegraphics[width=1.\columnwidth]{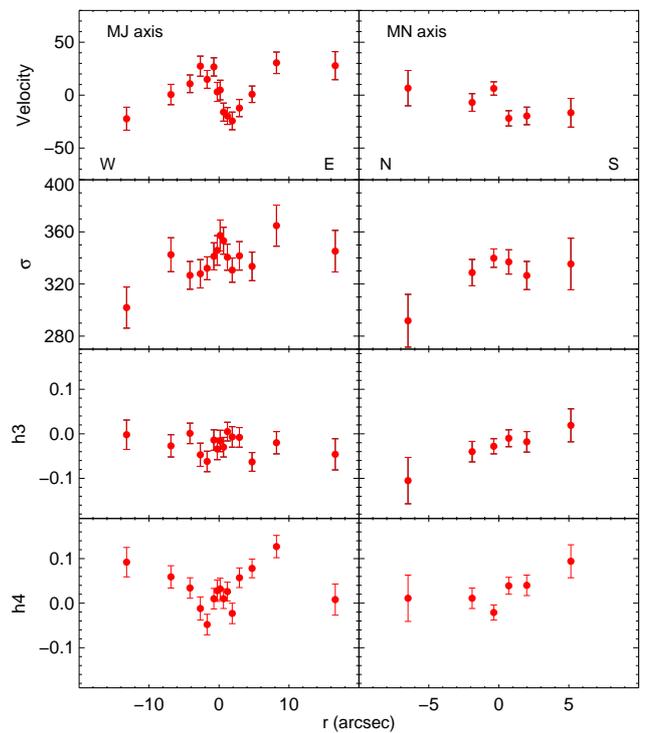}
\caption {Stellar kinematics along the major (MJ) and minor (MN) axes derived from the
  SALT data. Positive radii indicate projected distances E and S from
  the centre for the MJ and MN axes, respectively.}
\label{f_salt_kin}
\end{figure}

To recover the full LOSVD from the SALT data we used the Fourier
correlation method (FCQ, \citealt{Bender1990a}). While this method has the
advantage of minimizing template mismatch, since it operates in
Fourier space, the masking of spectral regions is
problematic. Therefore, to avoid masking the spectral regions of the
gap between the SALT detector chips, we derived the kinematics using
only the 5135$-$5431\AA\ range, where the strongest absorption features
are observed (e.g. Mgb, Fe5270). The SALT spectra were binned along
the spatial direction to achieve a minimum SNR of 30 per \AA\ at all
radii.  We used a 10~Gyr old synthetic stellar spectrum as kinematic
template \citep{Vazdekis2010}.  The uncertainties in the kinematic
parameters were estimated from Monte Carlo simulations of synthetic
spectra with artificial noise, based on the best-fitting set of
parameters for each spectrum.  Our stellar kinematic measurements
along the MJ and MN axes, are shown in Fig.~\ref{f_salt_kin}.

The kinematics derived from the SALT data in the inner $r \sim  
2$~arcsec is consistent with what is seen with \S: \n5419 shows little
rotation and a velocity dispersion of $\sim 350$\kms. In fact, while
the overall rotation amplitude is low, the long-slit data reveal that
the outer parts of the galaxy ($r>5$~arcsec) are rotating in the
opposite direction as the centre. The transition region, where the
sign of the angular momentum flips, is at $r \sim 5-6$~arcsec or
roughly four times the core radius $r_b$ (see Section~\ref{ss_core}).

\begin{figure*}
\centering
\includegraphics[width=0.77\textwidth]{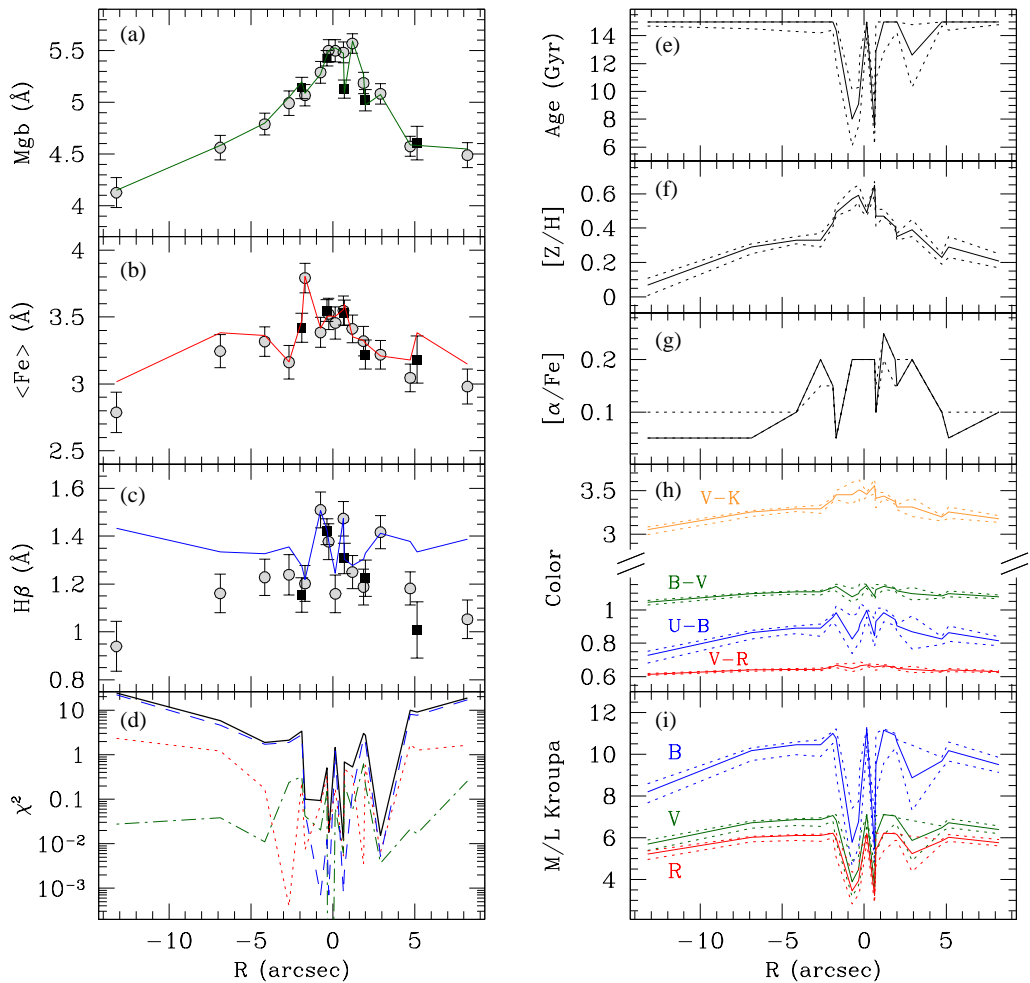}
\caption {Panels (a)$-$(c): measured Lick indices profiles along the major 
(circles) and minor (squares) axes. The lines correspond to the best-fit single 
stellar population models. Panel (d): $\chi^2$ for each line index (the 
green dot-dashed line corresponds to Mgb, the red dotted line to Fe and 
the blue dashed line to \Hb) and the total $\chi^2$ (black solid line).  
Panels (e)$-$(i): age, metallicity, overabundance, colours and M/L in different 
bands as a function of radius. The dotted lines show the 3$\sigma$ errors.}
\label{f_line_index}
\end{figure*}

\subsection{Line indices and stellar population analysis}\label{s_indices}

The wavelength range covered by SALT allows us to measure \Hb, Mgb5175, Fe5270 and Fe5335 line strength indices. These lines are known to be useful to constrain the age (particularly \Hb) and the chemical composition of the stellar population. Other lines in the SALT range that are commonly used to constrain stellar population parameters are Fe5015 and Fe5406. However, these lines are observed at the edges of the gaps between the detector chips (see Fig.~\ref{f_spectra}) and were not included in the following analysis. 

The Mg, Fe and \Hb\ line strength indices were derived for the binned spectra along the MJ and MN axis using the same synthetic star spectrum as for the kinematics. Its resolution was degraded to match that of the Lick system, and the galaxy's velocity dispersions measured in the previous section were also taken into account. We used the band definitions of \citet{Worthey1994a}. 
In the following we indicate the average Iron index with $\langle {\rm Fe} \rangle = ({\rm Fe5270} + {\rm Fe5335})/2$. 
To asses how well our measurements agree with the Lick system, we measured the line strength in several SSP model spectra from \citet{Vazdekis1999} multiplied by the SALT instrumental response and compared them with their Lick indices values. The largest discrepancy was found for the Fe5335 index, in which case we found a difference of 0.76~\AA\ between our measurement and the value given by \citet{Vazdekis1999}. The measured indices were corrected by these systemic offsets, and are shown in Fig.~\ref{f_line_index}. 

We study the stellar population using the SSP models of the Lick line indices of \citet{Maraston1998,Maraston2005} with a Kroupa IMF. These models cover ages of up 15~Gyr, 
metallicities [Z/H] from $-2.25$ to $+0.67$ and overabundances [$\alpha$/Fe] from $-0.3$ to $+0.5$ \citep[][]{Thomas2003a}. 
We follow the procedure of 
\citealt{Saglia2010a} (see their Section~4.1). At each radius, the age, metallicity and overabundance were derived by fitting the SSP models to the three line indices. 
Fig.~\ref{f_line_index} shows the resulting age, metallicity and overabundance derived from the line indices. We also include colours and M/L ratios for different bands corresponding to the best-fit SSP models. 
The models fit well the measured indices except for the case of \Hb, which is 
too low to be reproduced by the models, especially at larger radii. 
As a result, the age of the stellar population is found to be the maximum age explored by the  models, 15~Gyr (panel {\it e} of Fig.~\ref{f_line_index}). 
This is usually regarded as an effect of ionized gas emission, which partly fills the hydrogen absorption lines, leading to a weaken measurement of the \Hb\ index and, therefore, an  overestimation of the age of the stellar population. We do not see any evidence of gas emission in our spectra. However, \citet{Macchetto1996} reported the detection of weak H$\alpha + $\NII\ emission in the inner few arcsecond ($r\lesssim 5$~arcsec) of \n5419. 

Overall, the models point to an old, metal-rich, slightly $\alpha$/Fe-overabundant galaxy. The inferred $B-V$, $U-B$ and $V-R$ colours agree
reasonable well with those reported from broad-band photometry
\citep[$B-V=1.08$, $U-B=0.68$ and $V-R=0.67$ for
$r=11.4$~arcsec;][]{Poulain1994}, while the $V-K$ colour is similar to, though slightly redder than (possibly due to the overestimation of the stellar age) the value of $\sim3$ measured from the $HST$
and \S\ data (Section~\ref{s_photo}).  The mass-to-light ratios are
around 9, 5 and 4~\Ms/\Ls\ for the $B-$, $V-$ and $R$-band,
respectively (assuming a Kroupa IMF).

\section{Dynamical modelling}\label{s_dyn}

\begin{figure*}
\centering
\includegraphics[width=0.78\textwidth]{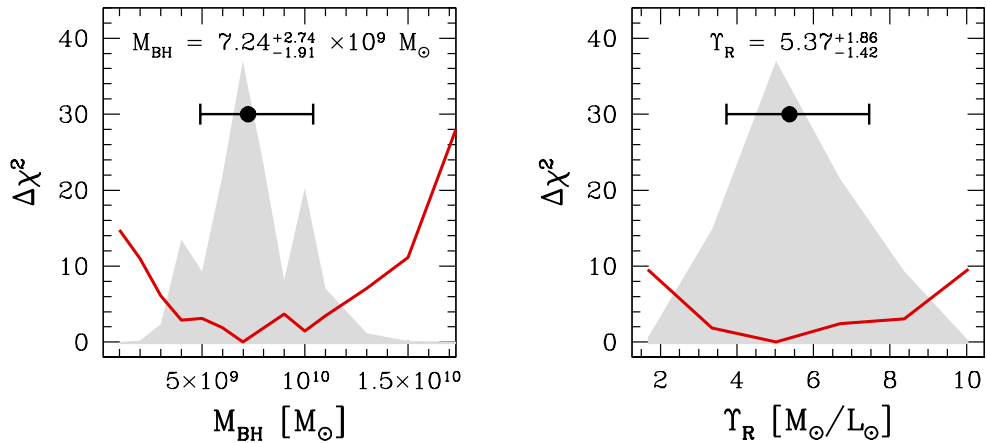}
\caption {The $\chi^2$ curves (solid-red line) and model likelihoods (grey;
  arbitrarily scaled) versus black hole mass $\mbh$ (left) and stellar
  mass-to-light ratio $\ml$ (right; extinction corrected). The
  one-dimensional model statistics were computed following
  \citet{McConnell2011b}. The point with error bars indicates the best-fit
  model and the 68 percent confidence interval.}
\label{f_chi2}
\end{figure*}

For the dynamical analysis of the \S\ and SALT spectra we assumed that
\n5419 is axisymmetric. The \S\ data consists of 29 symmetric
spatial bins in each quadrant. We first computed a folded kinematic 
data set by averaging the respective LOSVDs $\cal L$ of the four
quadrants. The LOSVDs measured via MPL were sampled at 23 bins in the
line-of-sight velocity $v_\mathrm{los}$. Each velocity channel was 
averaged individually, i.e. for a given $i \in [1,23]$ we averaged the
measured ${\cal L}_k (v_{\mathrm{los},i})$ ($k=1 \ldots 4$) of the
four quadrants, weighting each ${\cal L}_k$ by its error (and taking
into account that $v$ changes sign when crossing the galaxy's minor axis). We used these
folded SINFONI kinematics together with the long-slit data from SALT
as input for our dynamical models. The information contained in the wings 
of the LOSVDs is important for the estimation of the BH mass 
(e.g., \citealt{Nowak2010}, see also Appendix~\ref{s_appendix}). 

As discussed in Section~\ref{s_SINFkin}, there is an extended region of
high velocity dispersion in the innermost $\approx 100$~pc of \n5419
($\approx 0.35$~arcsec). It seems
morphologically connected to the double nucleus in the centre. As the
exact nature of this double nucleus is not clear at the moment (see
Section~\ref{s_nucleus}), we decided to omit all kinematical data points
inside $r<0.35$~arcsec. Accordingly, we used the photometric galaxy 
model discussed in Section~\ref{ss_core} as constraint for the light
distribution in the orbit model. In this way, our model is independent of whether 
the enhanced velocity dispersion near the centre is a 
feature of the global stellar population of \n5419 or whether 
it comes from a separate stellar component with a different orbital structure 
(associated with the double nuclear structure).

We constructed equilibrium models for \n5419 using our implementation of
Schwarzschild's orbit superposition method
\citep{Schwarzschild1979,Richstone1988,Gebhardt2003,Thomas2004,Thomas2005}. Schwarzschild
models are very flexible and do not require any a priori assumptions
upon the anisotropy of the stellar velocities. In brief, the model
construction requires four steps:
\begin{itemize}
\item Deprojection of the observed surface-brightness distribution to obtain the three dimensional intrinsic stellar luminosity density 
  $j$ \citep{Magorrian1999}.  

\item Setup of a trial mass distribution: 
\begin{equation}
 \rho = \ml \times j + \rho_\mathrm{DM} + \frac{\mbh}{4 \pi r^2} \times \delta(r), 
\end{equation}
assuming a stellar mass-to-light ratio $\ml$, a black hole mass \Mbh\ and a parametrised dark-matter (DM) halo density $\rho_\mathrm{DM}$ given by
\begin{equation}
 \rho_\mathrm{DM} = \frac{v_c^2}{4\pi G}~\frac{3r_c^2+r^2}{(r_c^2+r^2)^2},
\end{equation}
where $r_c$ is the core radius inside which the density slope of the DM is constant and $v_c$ is the asymptotic circular velocity of the DM. $\delta(r)$ denotes Dirac's delta function. 
\item Computation of a representative library of time-averaged stellar orbits.  
\item Solving for the set of orbital weights or occupation numbers, respectively, that minimises the $\chi^2$ difference between the model and the observed LOSVDs.  
\end{itemize} 
These four steps are repeated, varying independently the free
parameters of the model: $\ml$, \Mbh\ and the parameters of the dark
matter halo. The best-fitting \Mbh\ and $\ml$ and their errors are
derived from a one dimensional likelihood as in
\citet{McConnell2011b}. We used $\sim 27,000$ orbits to sample the
space of integrals of motion for each trial $\rho$. 
For the dynamical fits we used 125 constraints from the light distribution 
and 
759  
from the kinematic observations in the radial range $0.33 \le r \le
16.6$~arcsec. As noted above, we avoided the innermost LOSVDs because of the asymmetric
structure visible in the \S\ maps and in the \HST\ image.

Our best-fit model has $\mbh = 7.24^{+2.74}_{-1.91} \times 10^9$~\Ms\
and an extinction-corrected \Rb\ mass-to-light ratio of $\ml = 5.37^{+1.86}_{-1.42}$ (see
Fig.~\ref{f_chi2}). Models without black hole and/or without DM halo are 
ruled out by a $\Delta\chi^2 > 20$ and 40, respectively. 
The evidence for the BH comes from the high velocity wings of the LOSVDs 
inside the BH's sphere of influence and the constraints on the orbit 
distribution at larger radii \citep[][see also 
Appendix~\ref{s_appendix}]{Nowak2010,Rusli2013a}.
From the cumulative stellar mass distribution, we found a sphere-of-influence
radius of \n5419's black hole $-$ i.e. the radius at which the enclosed mass in stars equals the black hole mass $-$ of $r_\mathrm{SOI}
= 1.4$~arcsec. Thus, even when leaving out the very central data
points, our measurements resolve the central black hole's sphere of influence by more than a
factor of 4.

\begin{figure*}
\centering
\includegraphics[width=0.77\textwidth]{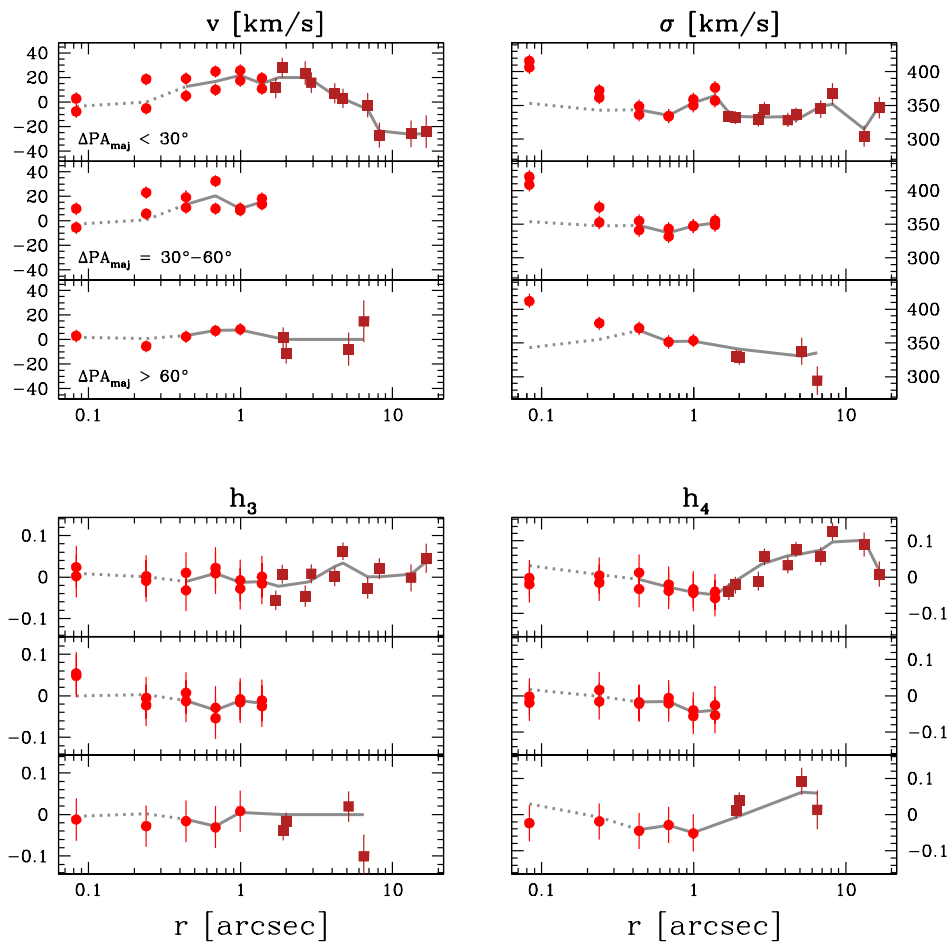}
\caption {Kinematic data and the best-fit model. Points with error
  bars show the SINFONI (red circles) and SALT (dark-red squares) folded kinematic data $-$ from
  top-left to bottom-right: rotation velocity $v$, dispersion
  $\sigma$, and Gauss-Hermite parameters $h_3$ and $h_4$. Within each
  panel, data points are separated by the position angle ($\Delta
  \mathrm{PA}_\mathrm{maj}$) relative to the galaxy's major axis. The grey solid
  lines indicate the best-fit dynamical model.  Data points inside
  $r<0.35$~arcsec were omitted in the fit (see Section~\ref{s_dyn}). The
  dotted lines show the inward extrapolation of the dynamical model
  down to the smallest measured radius. Note that for the dynamical   
modelling we use the full LOSVDs (Section~\ref{s_dyn}), which is essential for  
constraining the BH mass (see discussion in Appendix~\ref{s_appendix}).}
\label{f_dynmodel}
\end{figure*}

Fig.~\ref{f_dynmodel} shows the kinematics of the best-fit model
together with the (folded) data points. It also shows the innermost
data points that we omitted in the fit and the inward extrapolation of
the model into these regions. The model does not follow the strong
increase in $\sigma$ inside $0.35$~arcsec. This is not surprising. The
model is based on the fitted isophotes with the two nuclei in the
centre masked or subtracted, respectively. The flat central core in
the remaining light profile provides very little light in the centre
that could cause a strong gradient. The flat $\sigma$ distribution of
the model near the centre, despite the presence of a massive black hole implied 
by the fit, together with the morphological connection of the
high-$\sigma$ region and the double nucleus suggests that the high
dispersion near the centre comes from the extra light of one or both
of the two nuclei. Without higher spatial resolution data, we can not improve
our model in the innermost regions.

\subsection{SMBH and core properties}\label{s_BH-Core}

Since \n5419 is a core galaxy, we can investigate how well its
SMBH correlates with the core properties. The strongest correlation
found by \citet{Rusli2013b} was that between SMBH mass and core radius.
The core radius of $r_b=1.58$~arcsec ($\approx 430$~pc) that we obtained
from the fitting of a Core-S\'ersic function to the surface brightness
profile implies a black hole mass of $\mbh \approx 6.4 \times 10^9 \,
\msun$, in good agreement with our measurement. In other similar core 
galaxies a tight correlation between $r_{\rm SOI}$ and $r_b$ has been found,  
with $r_{\rm SOI} = r_b$ \citep{Thomas2016}. The $r_\mathrm{SOI}
= 1.4$~arcsec and $r_b=1.58$~arcsec of \n5419 are consistent with this 
relation.

With the mass-to-light ratio given by the dynamical models and the
light deficit estimated in Section~\ref{s_photo} ($\Delta L_R = 3.78 
\times 10^9 $~\Ls), we can also compute the stellar mass deficit in the
core of \n5419. This is simply given by $M_{\rm def} = \ml \times
\Delta L = 2.0 \times 10^{10}$~\Ms. The ratio between the mass deficit
in the core and the black hole mass, $M_{\rm def}/\mbh=2.8$, is well
within the range typically derived from observations and consistent
with the theoretical predictions from simulations of single dry
mergers \citep[see][and references therein]{Rusli2013b}.

\section{The double nucleus in \n5419 and high dispersion regions}\label{s_nucleus}

In Section~\ref{s_photo} we showed that the second nucleus observed
previously in {\it HST} images is also seen in our \S\ data. This
structure, located at 0.25~arcsec ($\sim 70$~pc) from the unresolved
central source, is relatively blue compared to the rest of the galaxy;
we obtained $V-K=1.68$ for the off-centre nucleus and $\sim 3$ for the
galaxy. Our \S\ velocity dispersion map indicates a central region of
high velocity dispersion ($\sigma \gtrsim 400$\kms) that includes both
nuclei and has a similar shape as the {\it HST} isophotes at
$r \sim 0.35$~arcsec (Fig.~\ref{f_sinf_kin_zoom}). In
Section~\ref{s_WCO} we ruled out the possibility that the steep
central $\sigma$ gradient connected with the two nuclei could be
explained as an artefact related to non-stellar continuum emission
associated with the LLAGN of \n5419. Thus, the high velocity
dispersion near the centre likely has a dynamical origin. We modelled
the galaxy assuming that the stars follow a Core-S\'ersic surface
brightness distribution, as indicated by the photometry at
$r\gtrsim 0.4$~arcsec (Sections~\ref{s_photo} and \ref{s_dyn}). The
inward extrapolation of our best-fit dynamical model does not,
however, explain the excess central velocity dispersion as the result
of the massive central black hole and the stellar orbits detected in
the (cored) main body of \n5419.  All this suggest that the high dispersion 
comes from a distinct stellar subsystem associated with the extra light 
in the double-nucleus region which we have omitted from the models.

The colour difference between the two nuclear light peaks in \n5419
makes it unlikely that the eccentric, elongated high-velocity
region is an asymmetric disk around a single black hole \citep[similar
to, though much larger, than the nuclear disk in M31;
e.g.][]{Tremaine1995,Bender2005}. Instead we assume that the
high-$\sigma$ stars originally formed a bound, undisturbed system
around one of the nuclei and that the present elongated structure is
the result of the two nuclei being in dynamical interaction. We can get an
order-of-magnitude mass estimate for the original component from
$\sigma^2={\rm G}M_{\star}/r_{\star}$, where we assume that
$r_{\star}$ is some 10~pc (upper limit from the larger of the two
nuclear structures, N2; cf. Section~\ref{s_photo}). Together with the observed
$\sigma = 400$\kms\, this points to a mass of about $M \sim 10^9\,\msun$,
with an uncertainty probably as large as a factor of $\sim 10$. 

The light emitted by the off-centre nucleus N2 only accounts for a
stellar mass of $\la 10^7$~\Ms\ (see Section~\ref{ss_massnuc}). If the
compact nuclear structure was originally associated with N2, then it
immediately follows that N2 hosts a SMBH with a mass of the order of
$10^9$~\Ms. Because of the lack of any gradient in the line-of-sight velocity 
of the high-$\sigma$ stars, however, it seems more plausible that they
were originally connected with the massive black hole at the galaxy's
photocentre (i.e., at rest relative to the main body of \n5419). This
would imply that a fraction of the spatially unresolved emission of N1
may still come from a very compact stellar component, in addition to
the galaxy's AGN.  The fact that the interaction with N2 could (partly)
dissolve this highly bound structure implies that N1 and N2 have
roughly similar masses. In conclusion then, irrespective of whether
the high-$\sigma$ stars originally belonged to N1 or N2, 
the data presented here strongly suggest that there are two SMBHs of similar 
mass (to within a factor of $\sim 10$) in the nucleus of \n5419, separated
by a distance of only $\sim 70$~pc.

If the centre of \n5419 really hosts two SMBHs, then N2 is probably
the remain of a galaxy nucleus, which is also suggested by the fact
that its colour and brightness are atypical for a globular cluster
(Section~\ref{ss_massnuc}). The light profile of \n5419 doesn't show
obvious distortions that would indicate an ongoing merger, yet the
kinematically decoupled core of \n5419 provides evidence for a
(probably minor) merger in the galaxy's latest evolutionary
phase. Since these mergers can last several Gyr (e.g. \citealt{Boylan-Kolchin2008}), 
we may be witnessing the last phase of such a minor merger.
Note, however, that the lack of a systemic radial velocity difference
between N1 and N2 implies a rather straight relative motion parallel
to the plane of the sky, while on larger scales the observed counter rotation requires
that at least some of the merger's orbital velocity was perpendicular
to the plane of the sky. 

If \n5419 is not in a late merger phase, could the two black holes 
have formed a stable, bound binary long ago? This is unlikely, since (for
two circularly orbiting similar SMBHs) one would expect both black
holes to be displaced from the centre of mass, i.e. the photocentre of
the galaxy.  Furthermore, the rather straight tidal feature seen between N1 and N2 
and the lack of a gradient in the line-of-sight velocity 
would imply that the binary orbit has to be of low angular momentum and we must be
seeing this binary very close to face-on, which is statistically
unlikely.

A more exotic explanation for two black holes not being in a close
orbit would be that we see a SMBH which was ejected from the centre on
a low angular momentum orbit in a past multiple merger event and is
now coming back from large radii.  While this scenario is
intrinsically unlikely too, the fast fly-by of N2 would
account for a number of observational aspects of \n5419. Firstly, it
would explain why we only see a straight tidal feature between N2 and
the central SMBH. Secondly, it would explain the fact that we cannot
detect signs of an ongoing merger any more. 

  In summary, the increased velocity dispersion at $r \la
  0.35$~arcsec strongly suggests that \n5419 hosts two SMBHs in close
  proximity near the centre. However, we cannot reach a definitive
  picture for the origin of the perturber N2.  The stellar kinematics
  at $r \ga 0.35$~arcsec are consistent with a black hole mass of
  $\mbh = 7.24^{+2.74}_{-1.91} \times 10^9$~\Ms\ located at the
  galaxy's photocentre (Section~\ref{s_dyn}). That this $\mbh$ is only
  slightly above the black hole mass we had expected if the galaxy
  were similar to other massive core ellipticals without a secondary
  nucleus suggests that the systematic uncertainties on $\mbh$ related
  to N2 are small.  Detailed numerical simulations and observations at
  a higher spatial resolution are required to put stronger constraints
  on the innermost mass distribution and on the origin of the two
  nuclei.

\section{Summary and conclusions}\label{s_summary}

We have presented high-resolution \Kb\ \S/VLT IFS and optical
long-slit spectroscopic observations of the galaxy \n5419. The
kinematics derived from these data show that \n5419 is a
dispersion-dominated galaxy. The rotational velocity does not exceed
50\kms; although a clear rotational pattern is observed, revealing a
counter-rotating core in the inner few arcseconds. The velocity 
dispersion is about 350\kms and almost constant over the low-surface
brightness core. However, inside $0.35$~arcsec ($\approx 100$~pc), where the galaxy hosts
a double nucleus, it increases reaching values of 420-430\kms.

We use orbit-based dynamical models to model the stellar kinematics
outside the double nucleus in the centre. From this analysis we derive a
\Mbh\ of $7.24^{+2.74}_{-1.91} \times 10^9$~\Ms. This mass is consistent
with the large core radius ($r_b=1.58$~arcsec or $\approx 430$~pc) obtained
by fitting a Core-S\'ersic function to the surface brightness profile,
given the known correlation between core radius and SMBH mass 
\citep{Rusli2013b}. 

The \Rb\ mass-to-light ratio derived from the dynamical modelling,
$\ml = 5.37$, is about a factor $1.1-1.4$ larger than the one derived 
from the stellar population analysis in the inner few arcseconds, for 
which we assumed a Kroupa IMF. The dynamical $\ml$ thus lies between 
a Kroupa and Salpeter IMF.

We have also discussed different scenarios in which the observed 
properties of the double nucleus of \n5419 can be explained. While the
nature of the double nucleus in \n5419 is certainly puzzling, our
observations suggest that this galaxy might host two SMBHs in close
proximity.  If this is the case, \n5419 is a promising target to study
the interaction between SMBHs at the centre of galaxies, and their
formation and evolution.  More clues could be obtained from
milliarcsecond radio imaging and higher spatial-resolution spectroscopic
data, that would allow us to study in a resolved manner the structures
seen in the central 0.5~arcsec (135~pc) of \n5419.

\section*{Acknowledgments}
Some observations used in this paper were obtained from the South African Large Telescope (SALT) under proposals 2012-1-DC-003 and 2012-2-DC-001.

\bibliographystyle{mnras}

\newcommand{\noop}[1]{}

\appendix
\section[]{What is stellar dynamical evidence 
for a BH?}\label{s_appendix}

This appendix is the result of a request of the referee to
present `prima facie' evidence for the existence of a BH in \n5419, 
by showing the measured stellar kinematic second order moments
and the best fitting models with and without a BH.

It has been known for 34 years (\citealt{Binney1982}, hereafter 
BM82) that centrally (i.e. near the BH sphere of influence) increasing
stellar velocity dispersion profiles do not provide `prima facie'
evidence for BHs. This is different from the case of gaseous disks
rotating around a BH, where a Keplerian increase of circular
velocities near the sphere of influence is expected to mark the
presence of the central BH \citep[e.g.][]{Barth2016a}. The recent increased
popularity of the Jeans approach to study dynamical masses in stellar systems 
\citep[e.g.][]{Tortora2009,Cappellari2013b,Newman2013,Feldmeier2013,Rys2014,
Scott2015,Lyubenova2016} stresses the importance
of modelling stellar second-order kinematic moments, but cannot
exploit the information in the full stellar line-of-sight velocity
distributions, which is the key to breaking the degeneracy between
the gravitational potential and anisotropy pointed out by \citeta{Binney1982}, 
as
discussed by \citet{Gerhard93}.

\citeta{Binney1982} showed that a centrally increasing stellar velocity
dispersion profile, such as that produced by a stellar system with a BH
and isotropic velocity dispersions for the stars, can also be produced
by a system \textit{without} a BH, if the velocity dispersions have
radial anisotropy that is centrally increasing. Here we reconsider this
case and add two further examples: a centrally decreasing velocity
dispersion profile and a system with a flat velocity dispersion.

These three (non-rotating) toy models are constructed via orbit superposition. 
In contrast to the usual application, where the orbital weights are
calculated so as to fit the (measured) projected kinematics as well as possible, here we  
constrain the intrinsic velocity moments of the orbit
superposition, and use the library to calculate the corresponding
LOSVDs. For the toy models, we assume a spherically symmetric stellar
distribution (using the spherically averaged luminosity density of \n5419) with 
a BH in its centre and the framework of \citet{Richardson2013}. For a 
given stellar mass, BH mass and 
anisotropy profiles $\beta(r)$ and $\beta^{\prime}(r)$ in the 2nd and 4th
order moments, respectively, we solve the Jeans Equations up to 4th
order. We calculate an orbit library similar to the ones used to fit
\n5419 but here we calculate the orbital weights such that the
respective velocity moments of the orbit superposition reproduce the
solutions of the Jeans Equations. We then calculate the LOSVDs of the
toy model using the same spatial bins on the sky as for \n5419.  We
have tested the method with the analytic (isotropic) models of 
\citet{Baes2005}. For an isotropic Hernquist sphere with
various BH masses (as studied in \citealt{Baes2005}) we find only slight
differences, and only in the higher-order moments ($\delta H_4 <
0.02$). We have constructed three different toy models, each of which
assumes a stellar $\Upsilon = 10$: (1) an isotropic model without a BH
($\beta = \beta^{\prime} \equiv 0$); (2) an isotropic model 
with a BH of mass $M_\mathrm{BH} = 7 \times 10^9 M_\odot$; (3) a model 
with the same BH mass but with a tangentially anisotropic orbit
distribution ($\beta = \beta^{\prime} < 0$). The tangential anisotropy
was chosen such that the model has a roughly constant central $v_{\rm rms}$.

\begin{figure*}
\centering
\includegraphics[width=0.9\textwidth]{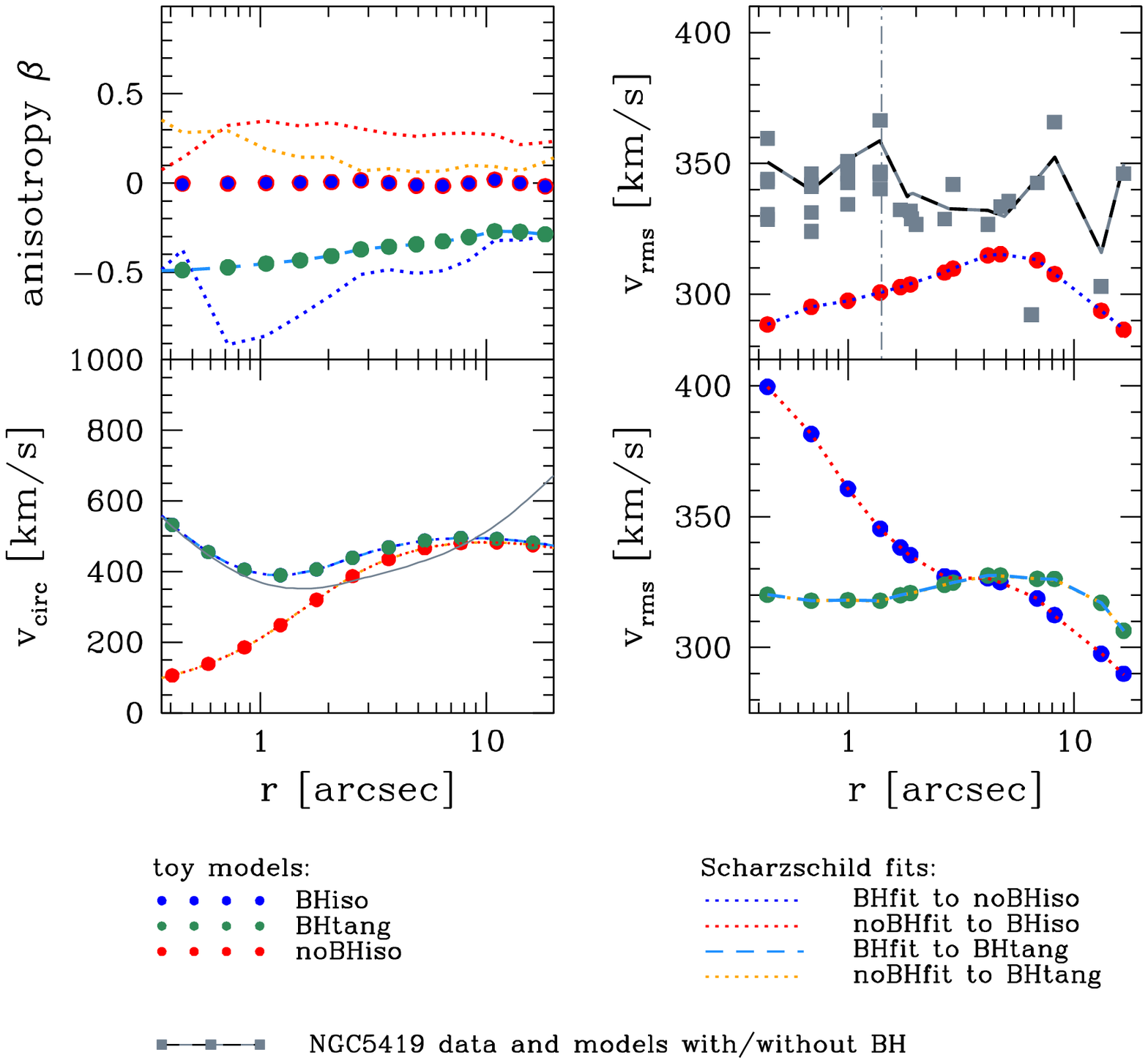}
\caption{Anisotropy profiles (top-left panel), circular velocity 
curves (lower-left panel) and projected 2nd order moments $v_\mathrm{rms}$ 
(right panels). Dots show the values from the toy models while dotted and 
dashed lines show the fits to the model data. Also shown are the data (grey 
squares) and best-fitting Schwarzschild model with and without a BH obtained 
for \n5419 (solid grey and dashed black lines). The vertical dot-dashed line 
indicates the $r_{\rm SOI}$ estimated for \n5419's BH (see 
Section~\ref{s_dyn}). BHiso, noBHiso and BHtang indicate the isotropic toy 
models with and without BH and the toy model with BH and tangential anisotropy, 
respectively. noBHfit and BHfit indicate the fits done assuming no BH and a 
$M_\mathrm{BH} = 7 \times 10^9 M_\odot$, respectively. 
For the toy models and \n5419, 
fits with or without a BH do not produce noticeable differences in the 2nd 
order projected velocity moment.}
\label{f_app-1}
\end{figure*}

The projected $v_{\rm rms}$ in the isotropic model with (without) 
a BH increases (decreases), as expected (Fig.~\ref{f_app-1}). Although this has
sometimes been taken to be the `self-evident' behaviour of a system
with (without) a BH, it reflects the assumed distribution of orbits in phase 
space as much as it reflects the underlying mass distribution.  
To illustrate this, we have fitted the toy models assuming: (1) a 
$M_\mathrm{BH} = 7 \times 10^9 M_\odot$ for the case of the toy model without a 
BH; (2) no BH in the case of the isotropic toy model with a BH and (3)  
both no BH and a $M_\mathrm{BH} = 7 \times 10^9 M_\odot$ in the case of the 
toy model with flat $v_{\rm rms}$.  
Each fit reproduces the 2nd moments of the input LOSVDs perfectly
(Fig.~\ref{f_app-1}). A dispersion profile that rises towards the
centre can reflect either an isotropic system with a concentrated mass
distribution, or a radially anisotropic system without a BH (see
Fig.~\ref{f_app-1} and e.g. \citeta{Binney1982},
\citealt{Binney2008}).  Conversely, a decreasing velocity dispersion
can be due to the lack of a central BH in an isotropic system (with
shallow inner mass distribution), or it can indicate tangential
anisotropy in such a system with BH. These results are expected to
depend on the spatial coverage and resolution of the kinematic data,
and on the luminosity constraints for the model.

\begin{figure*}
\centering
\includegraphics[width=0.78\textwidth]{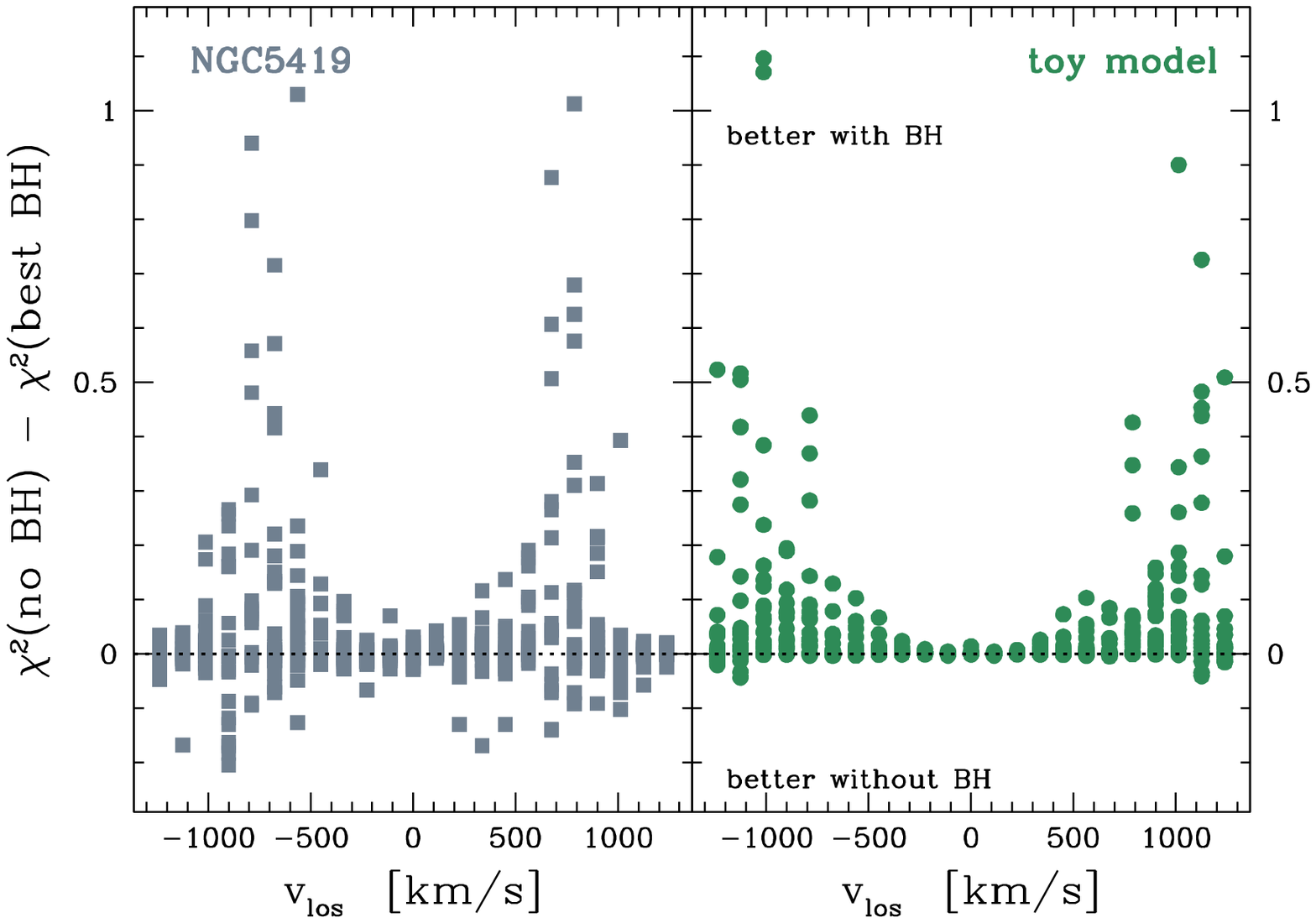}
\caption{Differences between best-fitting models with and without a BH for
\n5419 (left) and the toy model with a constant central velocity
dispersion (right). Each dot shows the $\chi^2$ difference in a single
line-of-sight velocity bin, and the differences are shown over all
LOSVDs within the sphere-of-influence radius of the best-fitting BH. 
The $\chi^2$ is normalised to the same rms as in \n5419. }
\label{f_app-2}
\end{figure*}

The 2nd moments provide only limited information about the actual
LOSVDs. For the toy model with nearly constant central dispersion,
we compare how well the fits with and without a BH reproduce the
input LOSVDs in Fig.~\ref{f_app-2}. Despite the fact that the fit without a BH 
{\it does} reproduce the 2nd moments of the LOSVDs, it clearly does {\it not} 
reproduce the LOSVDs as a whole. The residuals mainly occur at large projected
velocities, in the wings of the LOSVDs, and resemble the differences
between our best fitting models with and without a BH for \n5419
(Fig.~\ref{f_app-2}).

The Schwarzschild models are fitted to the entire LOSVDs, rather than
to a set of velocity moments. In our fits (for both the toy models as
well as for \n5419), we therefore minimise the difference between
the observed and modelled LOSVDs. The Schwarzschild models do not obey
any restrictions in the velocity moments, but have the freedom to
explore the full range of anisotropies in the moments of any order and
in the mixed terms of the velocity moment tensors that are allowed by
the tracer density distribution and gravitational potential. This
represents the most conservative approach to extract information about
the mass structure of an observed galaxy, as the models have the full
freedom to utilise all velocity moments to minimise the difference
between model and data. Residuals in the fits, as found here between
the models with and without a BH for \n5419, provide significant
evidence for the BH in the galaxy.

\bsp	
\label{lastpage}
\end{document}